\documentclass[twocolumn,showpacs,showkeys,preprintnumbers,amsmath,amssymb]{revtex4}
\usepackage{graphicx}
\usepackage{dcolumn}
\usepackage{bm}
\usepackage{subfigure}
\usepackage{caption}
\usepackage{ulem}
\begin{document}

\preprint{Draft for PRE}

\title{Hurst Exponents For Short Time Series\\}
\author{Jingchao Qi}
\author{Huijie Yang}
\email{hjyang@ustc.edu.cn} \altaffiliation {Corresponding author}
\address{Business School, University of Shanghai for Science and Technology,Shanghai 200093,China\\}
\date{\today}

\begin{abstract}
A new concept, called balanced estimator of diffusion entropy, is
proposed to detect scalings in short time series. The effectiveness
of the method is verified by means of a large number of artificial
fractional Brownian motions. It is used also to detect scaling
properties and structural breaks in stock price series of Shanghai
Stock market.

\end{abstract}

\maketitle

\section{Introduction}
\label{intro}

Scale invariance has been found to empirically hold for a number of
complex systems \citep{stanley01}. Consider a stochastic trajectory
$X(t)$, whose statistical properties are described by the
probability distribution function (PDF) of the displacements,
$p(x,t)$. The stochastic process represented by $X(t)$ behaves
scale-invariant, provides the PDF satisfies,
\begin{equation}
p(x,t)=\frac{1}{t^{\delta}}F\left(\frac{x}{t^{\delta}} \right),
\end{equation}
\noindent where $\delta$ is the scaling exponent. Ordinary
statistical mechanics is intimately related to the Central Limit
Theorem \citep{CLT49}, which implies the Gaussian form of the
function $F(\cdot)$ and $\delta=0.5$ \citep{Golden85}. By using the
the scaling exponents one can describe quantitatively the deviations
from ordinary mechanics, and consequently assess the real physical
nature of a phenomenon. But evaluation of the scaling exponents
meets several challenges.

How to evaluate a reliable scaling is not a trivial task.
Variance-based methods are used widely in literature to calculate
the scaling exponents\citep{Variance77} , in which there exists an
assumption of the time dependence of the variance $Var(X(t))$ with
the scaling exponent $\delta$, namely, $Var(X(t))\sim t^{2\delta}$.
For fractional Brownian motions it is valid, but there are scaling
processes such as L\'{e}vy flights for which the second moment
diverges or L\'{e}vy walks for which the second moment satisfies a
scaling relation $Var(X(t))\sim t^{2H}$ with $H\ne \delta$, i.e, the
relationships are violated \citep{Shles87}. Several efforts have
been done to develop complementary methods to evaluate reliable
scaling exponents \citep{Lafa01,Lafa10}. To cite an example, from
the PDF one can calculate the Shannon entropy, $S(t) = - \int
{p(x,t)\ln p(x,t)dx}$, which is originally identified as diffusion
entropy by Scafetta \textit{et al.} \citep{Lafa01}. It is proved
that the diffusion entropy can provide simultaneously reliable
values of $\delta$ for fractional Brownian motions and Levy
processes.

For a real-world stochastic process, the PDF is generally not known.
One can count how often the value $x$ appears in the data set of
trajectory $X(t)$. Denoting the number with $n(x)$, the PDF can be
estimated with the relative frequency, $\frac{n(x)}{N}$. $N$ is the
total size of the data set. In many situations only small data sets
from which to infer PDF are available. What is more, for a
stochastic process with a large amount of data sets, there exist
generally structural breaks in the trajectory due to emergent
strikes from environments and/or the system's transition to a
contrasting dynamical regime. To cite examples, a stock market is
shocked frequently by currency and tax policies, before and after an
earthquake the earth motion may stay in different dynamical regimes.
We should separate the data sets into many sub-sets to detect the
behaviors at different structural patterns. A small value of $N$ may
induce large statistical fluctuations or even bias to physical
quantities as the PDF, entropy, moments and so on. In a recent novel
paper, Bonachela \textit{et al.} recall the search for improved
estimators of entropy for small data sets \citep{Bonac08}. They
propose also a new "balanced estimator" that out-performs other
currently available ones when the data sets are small and $p(x,t)$
are not close to zero.

Stimulated by the two mentioned efforts, in the present paper a new
concept is introduced, called Balanced Estimator of Diffusion
Entropy (BEDE), in which the balanced estimator of entropy is used
to replace the original form in the diffusion entropy. This concept
is used to find scalings and structural breaks in artificial and
empirical series. Firstly, we review briefly the concepts of
diffusion entropy and balanced estimator, and introduce consequently
the concept of BEDE. Secondly, the effectiveness of the BEDE in
detecting scalings in short time series are verified by means of a
large amount of fractional Brownian motions. Finally, we detect the
scalings and structural breaks in the stock price series of Shanghai
Stock Market.

\section{Methods and Materials}
To keep the description self-contained, we review briefly the
concepts of diffusion entropy and balanced estimator.
\subsection{Diffusion Entropy}
Let us consider a one-dimensional stationary time series,
\begin{equation}
\xi_1,\xi_2,\cdots,\xi_N.
\end{equation}
\noindent $N$ is the length of the series. All the possible segments
with length $s$ read,
\begin{equation}
X_i= \{ \xi_i,\xi_{i+1},\cdots,\xi_{i+s-1} \}, i = 1,2,\cdots,N-s+1.
\end{equation}
\noindent Now we regard the length $s$ as time, the vector $X_i$ can
be regarded as a stochastic trajectory of a particle starting from
its initial position $X_i(0)=0$. By this way, the time series (2) is
mapped to an ensemble containing $N-s+1$ realizations of a
stochastic process. The displacements read,
\begin{equation}
x_i(s)=\sum\limits_{j=1}^{s}X_i(j)= {\sum\limits_{j=i}^{i+s-1}
\xi_j},i = 1,2,\cdots,N-s+1.
\end{equation}

One can divide the displacement interval where the particle appears
into $M(s)$ bins, and reckon the number of the particle's
occurrences in each bin at time $s$. We denote the numbers with
$N_j(s),j=1,2,\cdots,M(s)$. The PDF can be na\"{i}vely approximated
by the relative frequency,
\begin{equation}
p(j,s)\sim \hat{p}(j,s)=\frac{N_j(s)}{N-s+1}, j=1,2,\cdots,M(s).
\end{equation}
\noindent The entropy of the diffusion process is consequently
determined, which reads,
\begin{equation}
S_{DE}(s)\sim S_{DE}^{naive}(s)=-\sum\limits_{j=1}^{M(s)}
\hat{p}(j,s)ln[\hat{p}(j,s)].
\end{equation}
\noindent This entropy is based upon the diffusion process
constructed from the original series (2), for this reason is called
Diffusion Entropy (DE)  \citep{Lafa02}.

The key step in calculation of the DE is how to choose the size of
the bins, $\epsilon(s)$. The easiest way is to assume it to be a
fraction of the square root of the variance of the original series
(2) and independent of $s$.

Now we assume the time series behaves scale-invariance, namely,
$p(j,s)$ obeys the relation (1),
\begin{equation}
\begin{array}{l}
p(j,s)=\frac{1}{s^\delta}
F\left(\frac{x_{min}(s)+(j-0.5)\epsilon(s)}{s^{\delta}} \right),\\
j=1,2,\cdots,M(s),
\end{array}
\end{equation}
\noindent where $x_{min}(s)$ is the smallest value of displacement,
i.e., $x_{min}(s)=min[x_1(s),x_2(s),\cdots,x_N(s)]$. Let us plug
$Eq.(7)$ into $Eq.(6)$. A simple computation leads to,
\begin{equation}
S_{DE}(s)=A+\delta ln(s),
\end{equation}
\noindent where $A=\int\nolimits_{-\infty}^{+\infty} dy F(y)
ln[F(y)] $.

The simple relation of $Eq.(8)$ can be used to detect scalings in
time series. It is the first tool yielding the correct scaling in
both the Gauss and the L\'{e}vy statistics. For this reason, it
attracts special attentions from diverse research
fields\citep{Dentopy11}.

\subsection{Balanced Estimator For Diffusion Entropy}
From the relation $Eq.(5)$ we have the ensemble average,
\begin{equation}
\left\langle {\hat {p}(j,s)} \right\rangle = \frac{\left\langle {N_j
(s)} \right\rangle }{N - s + 1} = p(j,s).
\end{equation}
In other words, the frequencies $\hat {p}(j,s)$ approximate the
probabilities with certain statistical error (variance) but without
any systematic error (bias). The frequencies $\hat {p}(j,s)$ are
unbiased estimators of the probabilities $p(j,s)$.

However, there is an important difference between $S_{DE}(s)$ and
$S_{DE}^{naive}(s)$ in $Eq.(6)$ stemming from the non-linear nature
of the entropy functional. Defining an error variable, $\mu(j,s) =
\frac{\hat {p}(j,s) - p(j,s) }{p(j,s) }$, and replacing $p(j,s)$ in
$S_{DE}$ by its value in terms of $\mu(j,s)$ and $\hat{p}(j,s)$, a
straightforward algebraic leads \citep{Roust99},
\begin{equation}
S_{DE}(s)=S_{DE}^{naive}+\frac{M(s)-1}{2(N-s+1)}+\mathcal {O}(M(s)).
\end{equation}
\noindent The leading order of error, $\frac{M(s)-1}{2(N-s+1)}$, is
a significant error for small $N-s+1$ and vanishes only as
$(N-s)\rightarrow \infty$. Consequently, $S_{DE}^{naive}(s)$ is a
biased estimator of $S_{DE}(s)$, i.e., it deviates from the true
entropy not only statistically but also systematically.

An improved estimator of $S_{DE}(s)$ should reduce the bias or the
variance as possible, which can be formulated as follows. Defining
$S_{DE}[p(j,s)] \equiv -p(j,s)ln[p(j,s)]$, we have
$S_{DE}(s)=\sum\limits_{j=1}^{M(s)}S_{DE}[p(j,s)]$. We want to find
an estimator,
\begin{equation}
\hat{S}_{DE}(s)\equiv \sum\limits_{j=1}^{M(s)} \hat{S}_{DE}[n(j,s)],
\end{equation}
\noindent so that the bias,
\begin{equation}
\Delta^2_{bias}(s) = \left(\left\langle \hat{S}_{DE}(s)
\right\rangle-S_{DE}(s)\right)^2
\end{equation}
\noindent or the mean squared deviation
\begin{equation}
\Delta_{stat}^2(s) = \left\langle \left( \hat{S}_{DE}(s)-
\left\langle \hat{S}_{DE}(s) \right\rangle \right)^2 \right\rangle
\end{equation}
\noindent or a combination of both are as small as possible.

Ignoring correlations between the elements of the distribution,
$n(j,s),j=1,2,\cdots,M(s)$, the problem can be reduced to minimize
simultaneously the bias and the variance for each summand,
\begin{equation}
\Delta^2_{bias}[p(j,s)]= \left(\left\langle \hat{S}_{DE}[n(j,s)]
\right\rangle-S_{DE}[p(j,s)]\right)^2,
\end{equation}
\noindent and
\begin{equation}
\Delta_{stat}^2[p(j,s)] = \left\langle \left( \hat{S}_{DE}[n(j,s)]-
\left\langle \hat{S}_{DE}[n(j,s)] \right\rangle \right)^2
\right\rangle,
\end{equation}
\noindent where the probability $p(j,s)\in [0,1]$, and $n(j,s)\in
\{0,1,2,\cdots,N-s+1\}$ is binomially distributed. To balance the
errors, we minimize the average error over the whole range of
$p(j,s)\in[0,1]$,
\begin{equation}
\Delta^2(j,s)=\int_0^1 dp(j,s) \cdot w[p(j,s)]\cdot\left[
\Delta^2_{bias}(j,s)+\Delta^2_{stat}(j,s) \right],
\end{equation}
\noindent where $w[p(j,s)]$ is a suitable weight function that
specific problem depends. Without extra knowledge of the probability
values, one can consider a simple case of $w[p(j,s)]=1$. Inserting
$Eq.(14)$ and $Eq.(15)$ into $Eq.(16)$, the average error is given
by,
\begin{equation}
\begin{array}{l}
 \Delta ^2(j,s) = \\
 \int_0^1 {dp(j,s)} \left\{ {\sum\limits_{n(j,s) = 0}^{N - s + 1}
{P_{n(j,s)} \left[ {p(j,s)} \right]\hat {S}^2_{DE} [n(j,s)]} } \right. \\
 + S_{DE}^2 [p(j,s)] \\
 {\begin{array}{*{20}c}
 {\left. { - 2S_{DE} [p(j,s)]\left( {\sum\limits_{n(j,s) = 0}^{N - s + 1}
{P_{n(j,s)} \left[ {p(j,s)} \right]\hat {S}_{DE} [n(j,s)]} }
\right)}
\right\} } \hfill \\
\end{array} }, \\
 \end{array}
\end{equation}
\noindent where $P_{n(j,s)}[p(j,s)]$ is the binomial distribution,
\begin{equation}
\begin{array}{l}
 P_{n(j,s)} [p(j,s)] = \frac{[N - s + 1]!}{n(j,s)![N - s + 1 -
n(j,s)]!}\times \\
 {\begin{array}{*{20}c}
 {{\begin{array}{*{20}c}
 \hfill \\
\end{array} }} \hfill \\
\end{array} }{\begin{array}{*{20}c}
 \hfill \\
\end{array} }\left[ {p(j,s)} \right]^{n(j,s)} \cdot [1 - p(j,s)]^{N - s + 1
- n(j,s)}. \\
 \end{array}
\end{equation}

The goal is to determine all the numbers
$\hat{S}_{DE}[n(j,s)],j=1,2,\cdots,N-s+1$ which leads to the minima
of the average error. The necessary condition is that all the
partial derivatives vanish, i.e.,
\begin{equation}
\frac{\partial \Delta^2 (j,s)}{\partial \hat{S}_{DE}[n(j,s)]}=0,
n(j,s)=1,2,\cdots,N-s+1.
\end{equation}
\noindent Detailed computation leads to,
\begin{equation}
\begin{array}{l}
\hat{S}_{DE}[n(j,s)]\\
=(N-s+2)\int_{0}^{1}dp(j,s)P_{n(j,s)}[p(j,s)]\cdot S_{DE}[p(j,s)]\\
=\frac{n(j,s)+1}{N-s+3}\sum\limits_{k=n(j,s)+2}^{N-s+3} \frac{1}{k}.
\end{array}
\end{equation}

The final improved estimator of diffusion entropy reads,
\begin{equation}
\hat{S}_{DE}(s)=\frac{1}{N-s+3}\sum\limits_{j=1}^{M(s)}
[N_j(s)+1]\cdot \sum\limits_{k=N_j(s)+2}^{N-s+3} \frac{1}{k},
\end{equation}
\noindent called Balanced Estimator of Diffusion Entropy (BEDE) in
this paper.

\subsection{Materials}
\subsubsection{Fractional Brownian Motions}
Fractional Brownian motions (fBm) \citep{Brown96}are used to verify
the effectiveness of BEDE in detecting scalings in short time
series. A fBm is a continuous-time Gaussian process depending on the
Hurst parameter $0\le H \le 1$. The fBm is self-similar in
distribution and the variance of the increments is given by
$Var(fBm(t)-fBm(s)) \sim |t-s|^{2H}$. The program wfbm.m in
Matlab$^{\circledR}$ is used to generate the fBm series.

\subsubsection{Shanghai Stock Exchange Indices}
The empirical data are the time series of stock price indices from
the Shanghai Stock Exchange (SSE) \citep{SSE11}, the world's $5$th
largest stock market by market capitalization at US$2.7$ trillion as
of Dec 2010. The current exchange was established on November 26,
1990 and was in operation on December 19 of the same year. We
collect totally $134$ closed stock price series starting from the
end of the year $1995$ to the end of June, $2010$, in which the
numbers of the stocks distribute in the categories of industry,
business, real estate, public utility, and comprehension are
$64,27,12,12$ and $19$, respectively. We consider also the stock
price indices of the five categories from December 6,1994 to June
30, 2010. The SSE index series starts from December 19,1990 and ends
at June 30, 2010.

For a closed stock price series, $p(t)$, one can construct the
corresponding return series,
\begin{equation}
r(t)=\frac{log_a [p(t+\Delta t)]}{log_a[p(t)]}.
\end{equation}
\noindent In the calculations, $\Delta t$ is selected to be $5$,
i.e., weekly return ratio is considered.

\begin{figure}
\scalebox{0.80}[0.80]{\includegraphics{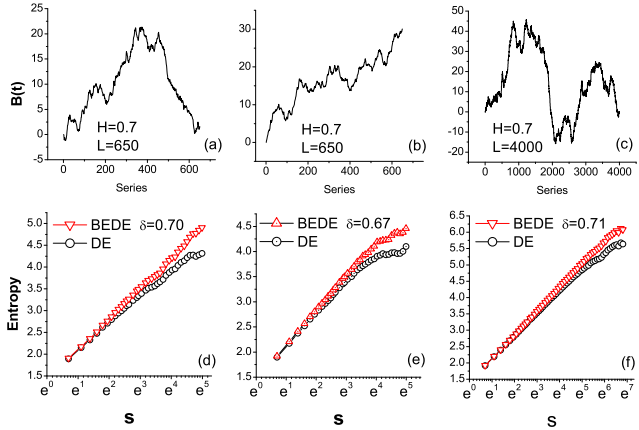}}
\caption{\label{fig:epsart} (Color online) Typical examples of
comparison between BEDE and DE. (a)-(c) Generated fBm motions with
$H=0.7$ and lengths $650,650,4000$, respectively. (d)-(f) In a wide
range of $s$, the BEDEs obey almost perfect linear relations versus
$ln(s)$. The estimated values of the scalings ( the slope of the
BEDE curve) are $0.70,0.67$ and $0.71$. With the increase of time,
the DE curves tend to bent down and the deviations from the linear
relations of $BEDE$ versus $ln(s)$ become more and more significant.
With the increase of length, e.g., $4000$, the DE curve is corrected
significantly to be much more closer to the BEDE curve in a wider
range of $ln(s)$, as shown in (f). }
\end{figure}

\begin{figure}
\scalebox{0.85}[0.85]{\includegraphics{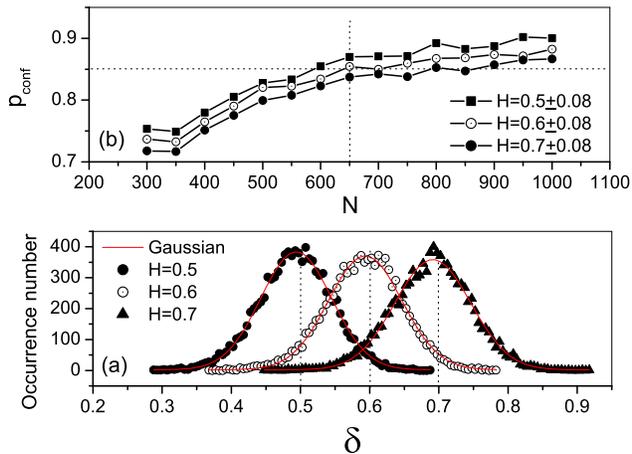}}
\caption{\label{fig:epsart} (Color online) Confidence of BEDE-based
scaling estimation. (a) $10^4$ series with $N=650$ and $H=0.5,0.6$
and $0.7$ are generated, respectively. The scaling estimations
distribute normally and center at the expected values of
$\delta=0.5,0.6$ and $0.7$. (b) The relation of certainty level
versus series length at a specified confidence interval $\Delta
H=0.08$. }
\end{figure}

\begin{figure}
\scalebox{0.85}[0.85]{\includegraphics{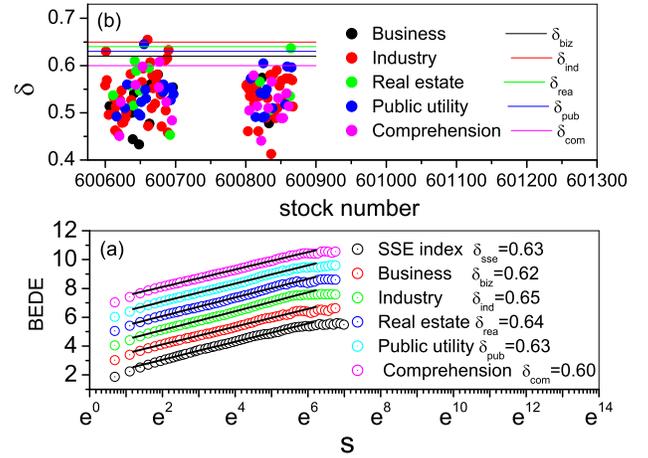}}
\caption{\label{fig:epsart} (Color online) BEDE-based scaling
estimations for the SSE index, and the indices of the five
catalogues including industry, business, real estate, public
utility, and comprehension. (a) In a wide range of $s$, the
relations of BEDE versus $ln(s)$ obey a linear-law. The scaling
exponents are
$\delta_{sse}=0.63,\delta_{biz}=0.62,\delta_{ind}=0.65,\delta_{rea}=0.64,\delta_{pub}=0.63$,
and $\delta_{com}=0.60$, respectively . (b) The scaling estimations
for the selected $134$ stocks. The scaling exponent for each
catalogue is significantly larger compared with that of the stocks
included in the corresponding catalogue. }
\end{figure}

\begin{figure}
\scalebox{0.85}[0.85]{\includegraphics{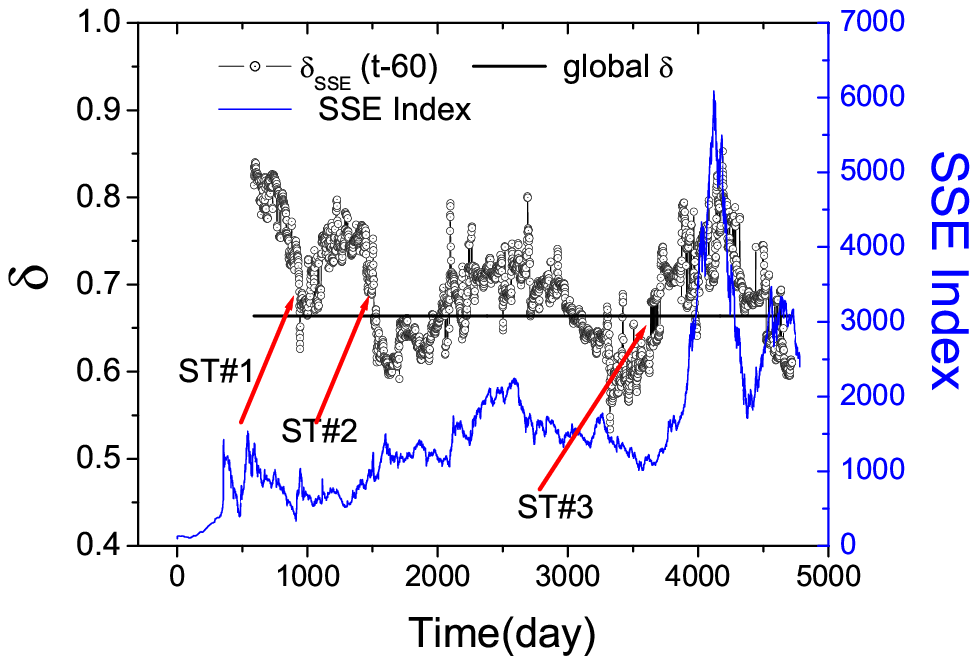}}
\caption{\label{fig:epsart} (Color online) Evolution of local
scaling estimation. Segment length is chosen to be $650$.  The SSE
index series and the $\delta$ evolution curve are matched with
$\Delta t=60$. The value of $\delta$ distributes in a wide interval
from $0.53$ to $0.85$. The scaling for the total series is $0.63$.
There exist globally four peaks covering $355,676,1704$ and $1396$
data points, namely, persisting roughly $18,34,85$ and $70$ months,
respectively. The significant transitions marked with ST\#1, ST\#2
and ST\#3 correspond to the bull market in the duration from July
29, 1994 to Sept. 13,1994, the bull market in the duration from
Jan.19,1996 to May 12,1997, and the RSSLC-induced increase of SSE
index for about 24 months (staring from May 9,2005), respectively.}
\end{figure}

\section{Results}
Fig.1 presents several typical examples of comparison between BEDE
and DE. For the three generated fBm series, (a)-(c), with $H=0.7$
and lengths $650,650,4000$, respectively, in a wide range of $s$,
the BEDEs obey almost perfect linear relations versus $ln(s)$ as
shown in (d)-(f), i.e., the scalings are all perfectly rendered out.
The estimated values of the scalings ( the slope of the BEDE curve)
are $0.70,0.67$ and $0.71$, which can be regarded as the same with
the expected value of $H=0.7$. However, for the short series in (a)
and (b), with the increase of time, the DE curves tend to bent down
and the deviations from the linear relations of $BEDE$ versus
$ln(s)$ become more and more significant. With the increase of
length, e.g., $4000$ in Fig.1(c), the DE curve is corrected
significantly to be much more closer to the BEDE curve in a wider
range of $ln(s)$, but at the region with larger values of $s$ it
still bents down with an unacceptable bias. Hence, for a series with
short length as $N\sim 650$, the calculated values of $DE$ have
unacceptable errors due to bias. The curve of $DE$ versus $ln(s)$
can not detect correctly the scaling at all, while the BEDE can give
perfect estimations of entropy even for considerable large of $s$,
namely, a small set of data ($N-s+1$ records). We must correct the
bias in entropy by means of the BEDE in analyzing the stock price
series of SSE, which has only a short history of about $15$ years
($\sim 10^3$ in length).

For a specific value of $H$, one can generate a large amount of fBm
series. It is found that the scaling estimations distribute
normally, as shown in Fig.2(a) a typical example, in which totaly
$10^4$ series with $N=650$ and $H=0.5,0.6$ and $0.7$ are used. By
specifying a confidence interval $[H-\Delta H, H+\Delta H]$ the
corresponding level of certainty $p_{conf}$ is determined so that
$p_{conf}\cdot N_{conf}$ estimations occur in the confidence
interval. $N_{conf}$ is the total number of the generated fBm
series. Fig.2(b) shows the relation of the certainty level
$p_{conf}$ versus the series length $N$. At the beginning, with the
increase of $N$, $p_{conf}$ increases rapidly, while when $N$
becomes large $p_{conf}$ tends to saturate to a high value.
Accordingly, we select $N=650$ in the following calculations in
detecting local scaling behaviors of the stock series in SSE stock
market.

We calculate the BEDEs for the Shanghai Stock Exchange index (SSE),
and the indices of the five catalogues including industry, business,
real estate, public utility, and comprehension. In a wide range of
$s$, the relations of BEDE versus $ln(s)$ obey a linear-law. Hence,
there exist almost perfect self-similarities and the scaling
exponents are
$\delta_{sse}=0.63,\delta_{biz}=0.62,\delta_{ind}=0.65,\delta_{rea}=0.64,\delta_{pub}=0.63$,
and $\delta_{com}=0.60$, respectively (see Fig.3(a)). The scaling
estimations for the selected $134$ stocks are also calculated, as
shown in Fig.3(b). The scaling exponent for each catalogue is
significantly larger compared with that of the stocks included in
the corresponding catalogue. Though each specific stock is almost
not predictable, the catalogue it belongs to, i.e., a combination of
the stocks in the catalogue is much more predictable.

For the SSE series, $\{r_{SSE}(1),r_{SSE}(2),\cdots,r_{SSE}(N)\}$,
one can calculate the scaling exponents for all the segments of
$\{r_{SSE}(t-s+1),r_{SSE}(t-s+2),\cdots,r_{SSE}(t)\},t=s,s+1,\cdots,N$,
denoted with $\delta_{SSE}(t-\Delta t)$, which are employed in the
present work to represent the local scalings of the SSE series, as
shown in Fig.4. The value of $s$ is chosen to be $650$. Assuming a
structural break occurs at time $t$, only when the segment covers a
certain number of data after the time $t$, the contribution from the
break's occurrence becomes significant and detectable. We introduce
the parameter $\Delta t$ to describe this kind of delay effect.

It is found that in the more than ten years duration the value of
$\delta$ distributes in a wide interval from $0.53$ to $0.85$. The
scaling for the total series is $0.63$, a value comparatively closer
to the lower bound $0.53$. What is more, though there exist rich
fine-structures with locally abrupt changes, there are globally four
peaks covering $355,676,1704$ and $1396$ data points, namely,
persisting roughly $18,34,84$ and $69$ months, respectively.

It is reasonable to believe that important events, such as policies
and/or emergencies, may lead speedy transitions of a stock market
from lower(higher) to higher(lower) predictable. From the
evolutionary curve of $\delta$, one can find three sharp transitions
marked with red arrows and denoted with $ST\#1£¬ST\#2$ and $ST\#3$,
respectively. The distances between the successive transitions are
about $17.5$ and $112.5$ months. By comparing with the important
events occurring in the history of the SSE market \citep{Modnews},
the value of $\Delta t$ is determined to be $60$. By this way the
stock series and the $\delta$ evolutionary series are matched along
time, as shown in Fig.4.

The first transition, $ST\#1$, corresponds to the bull market in the
duration from July 29, 1994 to Sept. 13,1994. Before this bull
market, the market has suffered from a $17$-month-duration of
decrease. The China Securities Regulatory Commission (CSRC) issues
three special policy regulation items to bailout the stock market.
Accordingly, the SSE index increases rapidly from $325$ to $1052$
within one and half month (reaches the record at Sept. 13,1994).

The second transition, $ST\#2$, matches with the bull market in the
duration from Jan.19,1996 to May 12,1997, in which the stock index
rises up to $1464$ from $512$. At the time speculating blue chip
stocks tends to dominate the investment concept. The Shenzhen
Development Bank and Sichuan Changhong become successively the
leading stocks in the Shenzhen Stock Exchange market and The
Shanghai Stock Exchange market, respectively. Stimulated by the two
stocks, the SSE market becomes high active and after the National
day of China, the prices for almost all stocks increase rapidly. The
CSRC issues successively some policy regulation items to cool down
the stock market and expounds in-detail the irrational state of the
stock market.

Starting from June 13,2001, the day a local maxima occurs at level
2242, a decreasing process persists about $48$ months, during which
a bouncing maxima occurs at level $1778$ at April 6, 2004. At May
9,2005 the reform of the shareholder structure of listed companies
(RSSLC) is conducted, which induces a persistent increase of SSE
index for about 24 months. This event accords with the third
transition, $ST\#3$. Then the persistent increase is disturbed to a
chaotic state by the escalation of stamp tax at May 30,2007.

\section{Conclusions}
Scaling invariance holds in a large amount of complex systems, but
the evaluation of scaling is still a challenge task. Theoretically,
variance-based methods can not detect correctly the scalings for
Levy processes. Empirically, time series are usually not long enough
to derive a reliable scaling exponent. What is more, in a long time
series there exist usually structural breaks. In literature,
diffusion entropy is developed to detect reliable scaling exponents
for long time series. In the present paper, the balanced estimation
of entropy for short time series is introduced to the diffusion
entropy to find reliable scalings embedded in short time series.

This method can give reliable scalings even for short time series
with length $\sim 10^2$. It is used to detect the scalings embedded
in totally 134 stocks in SSE market. The scaling exponent for each
catalogue is significantly larger compared with that for the
specific stocks included in this catalogue.

We detect also the local scalings in the SSE index series. The
scalings varies in a large interval from $0.53$ to $0.85$. Three
speedy transitions from high(low) to low(high) values of $\delta$
occur in the evolutionary curve of $\delta$, which are used to match
the $\delta$ curve with the closed price curve along time.

\begin{center}
\textbf{Acknowledgements}
\end{center}

The work is supported by the National Science Foundation of China
under Grant Nos. 10975099 and 10635040, the Program for Professor of
Special Appointment (Eastern Scholar) at Shanghai Institutions of
Higher Learning, and the Shanghai leading discipline project under
grant No.S30501.We thank the reviewers for their stimulating and
constructive comments and suggestions.

\end{document}